\documentclass[conference]{IEEEtran}%
\usepackage{amsfonts}
\usepackage{amssymb}
\usepackage{cite}
\usepackage{graphicx}
\usepackage[cmex10]{amsmath}
\usepackage{graphicx}

\IEEEoverridecommandlockouts

\ifCLASSINFOpdf
\else
\fi

\begin{document}
%
\title{Towards Energy Neutrality in Energy Harvesting Wireless Sensor Networks: A Case for Distributed Compressive Sensing?}
%
%
%

\author{
    \authorblockN{Wei Chen\authorrefmark{1}\authorrefmark{2},
        Yiannis Andreopoulos\authorrefmark{3},
        Ian J. Wassell\authorrefmark{2} and Miguel R. D. Rodrigues\authorrefmark{3}
        \thanks{This work is supported by EPSRC Research Grant EP/K033700/1 and EP/K033166/1, and the State Key Laboratory of Rail Traffic Control and Safety (RCS2012ZT014), Beijing Jiaotong University.}
        }
    \authorblockA{
        \authorrefmark{1}
        State Key Laboratory of Rail Traffic Control and Safety, Beijing Jiaotong University 100044, China}
    \authorblockA{
        \authorrefmark{2}
        Computer Laboratory, University of Cambridge, UK}
    \authorblockA{
        \authorrefmark{3}
        Department of Electronic \& Electrical Engineering, University College London, UK}
}

\maketitle

\begin{abstract}
This paper advocates the use of the emerging distributed compressive sensing (DCS) paradigm in
order to deploy energy harvesting (EH) wireless sensor networks (WSN) with practical network
lifetime and data gathering rates that are substantially higher than the state-of-the-art. In
particular, we argue that there are two fundamental mechanisms in an EH WSN: i) the energy
diversity associated with the EH process that entails that the harvested energy can vary from
sensor node to sensor node, and ii) the sensing diversity associated with the DCS process that
entails that the energy consumption can also vary across the sensor nodes without compromising
data recovery. We also argue that such mechanisms offer the means to match closely the energy
demand to the energy supply in order to unlock the possibility for energy-neutral WSNs that
leverage EH capability. A number of analytic and simulation results are presented in order to
illustrate the potential of the approach.
\end{abstract}




%
\IEEEpeerreviewmaketitle

\section{Introduction}
%
%
%
%
\IEEEPARstart{F}{uture} deployments of wireless sensor network (WSN) infrastructures are expected
to be equipped with energy harvesters (e.g. piezoelectric, thermal or photovoltaic) to
substantially increase their autonomy and
lifetime~\cite{kansal2007power,roundy2004power,sudevalayam2011energy}. However, it is also widely
recognized that the existing gap between the sensors' energy harvesting (EH) supply and the
sensors' energy demand is not likely to close in the near future due to limitations in current EH
technology, together with the surge in demand for more data-intensive
applications~\cite{sudevalayam2011energy}. Consequently, the realization of energy neutral (or
nearly energy neutral) WSNs for data-intensive applications remains a very challenging problem.

\par
These considerations have motivated the design of various emerging data acquisition and
transmission schemes and protocols for EH
WSNs~\cite{kansal2007power,sharma2010optimal,ozel2011transmission}. For example,
\cite{kansal2007power} proposes to characterize the complex time varying nature of energy sources
with analytically tractable models, \cite{sharma2010optimal} proposes energy management policies
that are throughput optimal and mean delay optimal, and \cite{ozel2011transmission} puts forth a
directional water-filling algorithm that maximizes the throughput by a deadline, and minimizes the
transmission completion time of the communication session. However, such existing energy
management approaches do not integrate organically fundamental mechanisms associated with the EH
process and the sensing process in an EH WSN: \emph{energy diversity} and \emph{sensing
diversity}.


\par
This article -- at the core of its contribution -- advocates the use of distributed compressive
sensing (DCS) in order to deploy WSNs with practical network lifetime and data gathering rates
that are substantially higher than the state-of-the-art. The key attributes of the proposed
approach that lead to efficient energy management are associated with the fact that -- subject to
certain conditions on the measurement process and the collection of signals to be sensed (e.g.,
intra- and inter-signal correlation) --\\
-- the number of data projections (measurements) at the various sensors can be substantially lower
than the data dimensionality without compromising data
recovery~\cite{candes2006robust,donoho2006compressed};\\
-- the number of data projections (measurements) at the various sensors can be adjusted without
compromising data recovery~\cite{baron2006distributed,duarte2011bounds}.\\
Since the number of projections acts as a proxy to energy efficiency -- in view of the fact that
transmission energy tends to be orders of magnitude higher than sensing/computational energy in
various WSNs applications~\cite{6155205,mamaghanian2011compressed} -- then the proposed approach
provides for i) substantial energy efficiency in relation to other approaches, such as methods
that do not exploit any form of source compression; and ii) adapting energy consumption to the
random nature of energy availability in EH systems.

\par
That is, the article argues that -- due to the energy diversity associated with the EH process
where the harvested energy can vary from sensor to sensor and the sensing diversity associated
with the DCS process where the number of projections can also vary from sensor to sensor -- one
should able to closely match the energy supply to the energy demand in order to unlock the
possibility for energy neutral operation in EH WSNs. This principle is illustrated in the sequel:
Section \uppercase\expandafter{\romannumeral2} describes in detail the proposed sensing approach.
Sections \uppercase\expandafter{\romannumeral3} and \uppercase\expandafter{\romannumeral4} present
analytic and simulation results that support the potential of the approach. General concluding
remarks are drawn in Section \uppercase\expandafter{\romannumeral5}.

\begin{figure}[t]%
\centering%
\includegraphics[width=0.30\textwidth]{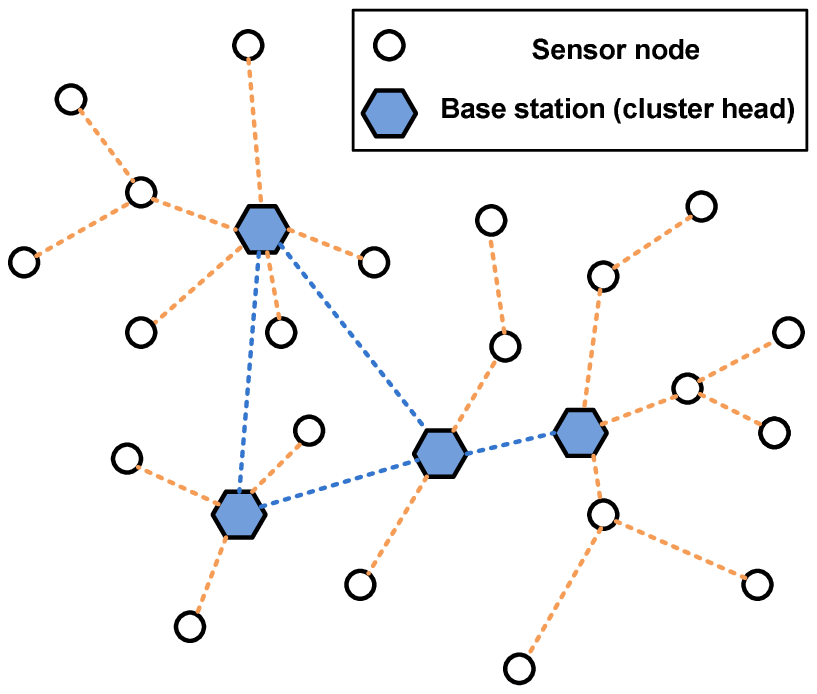}%
\DeclareGraphicsExtensions. \caption{A typical cluster-based WSN architecture.}
\label{fig:wsn-model}
\end{figure}%
\begin{figure}[t]%
\centering%
\includegraphics[width=0.35\textwidth]{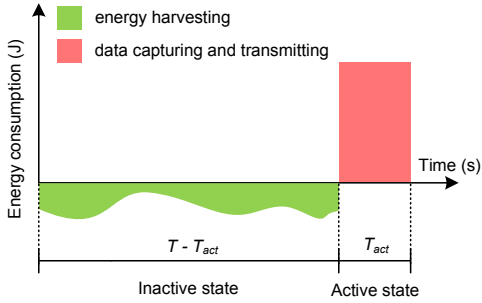}%
\DeclareGraphicsExtensions. \caption{Typical energy consumption profile of a data acquisition and
EH scheme.} \label{fig:energy-harvesting}
\end{figure}%

\section{System Description}
We consider a typical cluster-based WSN architecture, as shown in Fig.~\ref{fig:wsn-model}, where
a set of sensor nodes (SNs) periodically conveys data to one or more base stations (BSs). We
assume slotted transmission such that within a time slot of $T$ seconds the SNs are active for
$T_\text{act}$ seconds in order to capture and transmit data and are inactive for $T-T_\text{act}$
seconds in order to harvest energy from the environment, as shown in
Fig.~\ref{fig:energy-harvesting}. We also consider an innovative data gathering and reconstruction
process -- which is key to match the energy demand to the energy supply -- based on three key
steps: i) DCS based data acquisition, ii) transmission of projections, and iii) DCS based data
reconstruction.

\subsection{Data Acquisition}
The SNs capture low-dimensional projections of the original high-dimensional data during each
activation time $iT-T_\text{act}\leq t\leq iT$, which are given by:
\begin{equation}\label{eq:system_model1}
\mathbf{y}_k(i)=\mathbf{\Phi}_k(i) \mathbf{f}_k(i),
\end{equation}
where $\mathbf{y}_k(i)\in \mathbb{R}^{m_k(i)}$ is the projections vector at the $k$th SN
corresponding to the $i$th time interval, $\mathbf{f}_k(i)\in \mathbb{R}^{n(i)}$ is the original
(Nyquist-sampled) data vector at the $k$th SN corresponding to the $i$th time interval, and
$\mathbf{\Phi}_k(i) \in \mathbb{R}^{m_k(i)\times n(i)}$ is the projections matrix where $m_k(i)\ll
n(i)$ for any time interval $i$ and SN $k$. Note that the dimensionality of the projections can
vary in different activation times and different sensor nodes. In practice, one may obtain the
projections vector from the original data signal using analogue CS
encoders~\cite{6155205,mishali2011xampling}, whereby the projections vector is obtained directly
from the analogue continuous-time data, or using digital CS encoders~\cite{6155205}, whereby the
projections vector is obtained from the Nyquist sampled discrete-time data via
(\ref{eq:system_model1}). Recent studies suggest that digital CS encoders are more energy
efficient than analogue CS encoders for WSNs~\cite{6155205}.


\subsection{Data Transmission}
The SNs then convey the low-dimensional projections of the original high-dimensional data to the
respective fusion centers. We assume that upon activation the SNs converge into a balanced
time-frequency steady-state mode where each SN is associated with a BS using a particular channel
(or joins a synchronized channel hopping schedule) in order to convey data without collisions. We
also assume that fading, external interference noise and other non-idealities in packet
transmissions are dealt with via the PHY layer modulation and coding mechanisms of standards like
IEEE 802.15.4. Therefore, the transmission of the projections of the original data is taken to be
essentially perfect without a significant loss in generality.


\subsection{DCS Based Data Reconstruction}
We take the signals $\mathbf{f}_k(i)\in \mathbb{R}^{n(i)}$ to admit a sparse representation
$\mathbf{x}_k(i)\in \mathbb{R}^{n(i)}$ in some basis $\mathbf{\Psi}(i) \in \mathbb{R}^{n(i)\times
n(i)}$, i.e.
\begin{equation}\label{eq:system_model2}
\mathbf{f}_k(i)=\mathbf{\Psi}(i) \mathbf{x}_k(i),
\end{equation}
where $\|\mathbf{x}_k(i)\|_0=s_k(i)\ll m_k(i) \ll n(i)$. In addition, we also take the sparse
representations to obey the sparse common component and innovations (SCCI) model that has been
frequently used to capture intra- and inter-signal correlation typical of physical signals (e.g.,
temperature, humidity) in WSNs~\cite{baron2006distributed}, i.e.,
\begin{equation}\label{eq:SCCI_model}
\mathbf{x}_k(i)=\mathbf{z}_c(i)+\mathbf{z}_k(i),
\end{equation}
where $\mathbf{z}_c(i)\in \mathbb{R}^{n(i)}$ with $\|\mathbf{z}_c(i)\|_0=s_c'(i)\ll n(i)$ denotes
the common component of the sparse representation $x_k(i)\in \mathbb{R}^{n(i)}$, which is common
to the various SNs, and $z_k(i)\in \mathbb{R}^{n(i)}$ with $\|\mathbf{z}_k(i)\|_0=s_k'(i)\ll n(i)$
denotes the innovations component of the sparse representation $x_k(i)\in \mathbb{R}^{n(i)}$,
which is specific to each SN. Note that $s_c'(i)+s_k'(i)\geq s_k(i)$.

\par
The typical signal reconstruction process behind conventional CS approaches involves solving the
following optimization problem to recover individually the original signals captured by the
various sensors:
\begin{equation}\label{eq:CS_L1}
\begin{split}
\min_{\mathbf{x}_k(i)} \qquad &\|\mathbf{x}_k(i)\|_{1}\\
\text{s.t.} \qquad &\mathbf{A}_k(i)\mathbf{x}_k(i)=\mathbf{y}_k(i),
\end{split}
\end{equation}
where $\mathbf{A}_k(i)=\mathbf{\Phi}_k(i)\mathbf{\Psi}(i) \in \mathbb{R}^{m_k(i)\times n(i)}$.

\par
In contrast, the signal reconstruction process behind the adopted DCS approach involves solving
the following optimization problem to recover jointly the original signals captured by various
SNs:
\begin{equation}\label{eq:CS_DCS}
\begin{split}
\min_{\tilde{\mathbf{z}}(i)} \qquad &\|\tilde{\mathbf{z}}(i)\|_1\\
\text{s.t.} \qquad &\tilde{\mathbf{A}}(i)\tilde{\mathbf{z}}(i)=\tilde{\mathbf{y}}(i),
\end{split}
\end{equation}
where $\tilde{\mathbf{z}}(i)=\left[\mathbf{z}_c(i)^T\ \mathbf{z}_1(i)^T\ \ldots\
\mathbf{z}_K(i)^T\right]^T\in \mathbb{R}^{(K+1)n(i)}$ is the extended sparse signal vector,
$\tilde{\mathbf{y}}(i)=\left[\mathbf{y}_1(i)^T\ \ldots\ \mathbf{y}_K(i)^T\right]^T\in
\mathbb{R}^{\sum_{k=1}^K m_k(i)}$ is the extended measurements vector, and
$\tilde{\mathbf{A}}(i)\in \mathbb{R}^{\left(\sum_{k=1}^K m_k(i)\right) \times (K+1)n(i)}$ is the
extended sensing matrix given by
\[\tilde{\mathbf{A}}(i)=
\begin{bmatrix}
\mathbf{A}_1(i)&\mathbf{A}_1(i)&\mathbf{0}&\mathbf{0}  & \cdots & \mathbf{0}      \\
\mathbf{A}_2(i)&\mathbf{0}&\mathbf{A}_2(i)&\mathbf{0}  & \cdots & \mathbf{0}      \\
\vdots &  &    &                    & \ddots     & \vdots \\
\mathbf{A}_K(i)&\mathbf{0}&\mathbf{0}&\mathbf{0}    & \cdots & \mathbf{A}_K(i)
\end{bmatrix}.\]

\par
Note that the reconstruction procedure in (\ref{eq:CS_DCS}) -- in contrast to the reconstruction
procedure in (\ref{eq:CS_L1}) -- exploits not only intra- but also inter-signal correlation in
order to provide further efficiency gains that are leveraged by our DCS based energy management
scheme.

\subsection{Energy Consumption and Harvesting Models}
The data gathering process is subject to a causal energy harvesting constraint. In particular, we
assume that the SNs have to obey a certain energy budget during each activation interval
$iT-T_\text{act}\leq t\leq iT$, which is given by:
\begin{equation}\label{eq:energy_budget}
\xi_k(i)=\min\left\{\xi_k^\texttt{tot};\left(\sum_{j\leq
i}\xi_k^H(j)\right)-\left(\sum_{j<i}\xi_k^C(j)\right) \right\},
\end{equation}
where $\xi_k^\texttt{tot}$ is the capacity of the battery of SN $k$, $\xi_k^H(j)$ is the energy
harvested by SN $k$ in the interval $(j-1)T\leq t\leq jT-T_\text{act}$ and $\xi_k^C(j)$ is the
energy consumed by SN $k$ in the interval $jT-T_\text{act}\leq t\leq jT$. Note that this budget is
positive because the cumulative consumed energy has to be less than the cumulative harvested
energy.

\par
We adopt the following models that are relevant to determine the energy harvested and the energy
consumed in each activation interval. We assume that energy consumed for sensing, computing and
transmitting one measurement is a constant\footnote{This assumption is motivated by the fact that
if the computation is regular (which is the case in CS-based and and DCS-based data gathering) and
the PHY/MAC layers are not adapting the modulation, coding and retransmission strategies during
the active time (which is the case under low-energy IEEE $802.15.4$ PHY and MAC-layer processing
under SN-oriented operating systems -- e.g., nullMAC in the Contiki OS) then computing and
transmitting one measurement will come at quasi-constant energy consumption.} $\tau>0$. Hence, the
energy budget $\xi_k(i)$ for transmitting $m_k(i)$ measurements should satisfy:
\begin{equation}\label{eq:measurement_energy_relationship}
\xi_k(i)\geq \tau m_k(i).
\end{equation}

\par
We also assumed that the harvested power follows a uniform distribution in the interval
$[(1-\rho)\mu, (1+\rho)\mu]$, where $\mu$ denotes the mean harvested power and $0< \rho \leq 1$.
We adopt this particular simple model in view of the fact that power harvesting with current
technologies for solar and piezoelectric harvesters is either piecewise uniformly distributed (or
a mixture of two uniform probability density functions (PDFs) around the peak harvesting and
minimal harvesting values) or simply uniformly distributed over the range of power that can be
generated by the energy harvester~\cite{kansal2007power,roundy2004power,sudevalayam2011energy}.
For example, piezoelectric harvesting with a 1 $\texttt{cm}^2$ panel can be assumed to derive
power that is piecewise uniformly distributed between $[0, 200\mu \texttt{W}]$, or simply
uniformly distributed within this range~\cite{sudevalayam2011energy}.


\par
We also adopt throughout a very simple energy management approach where the SNs use the entire
available energy budget per activation interval, i.e., the number of projections per SN are such
that the energy consumption fits into the harvested energy budget (the unused energy is taken to
be lost from one activation interval to a subsequent one). Therefore, we drop the activation
interval index $i$ in the ensuing analysis to simplify the notation.


\section{Towards Energy Neutrality: an Analysis of the Two Sensor Scenario}
We illustrate the essence of the approach in an EH WSN consisting of two SNs, where the energy
harvested by each SN is independent. We compare the main attributes of the DCS data gathering
scheme to a CS and a raw data acquisition method.
\par
The simplicity of the two sensor scenario unveils the potential of the DCS approach to unlock
higher levels of energy efficiency in EH WSNs.

\subsection{Raw Data Gathering Scheme}
The probability of incorrect data reconstruction in a raw data gathering scheme can be lower
bounded by the probability that the energy harvested by the two SNs is not sufficient to fit the
energy consumption requirements by each of the two SNs. Hence, by using the assumption that the EH
processes follow independent uniform distributions in the interval $[(1-\rho)\mu, (1+\rho)\mu]$ it
follows that\footnote{We assume in this sub-section and in sub-sections
\uppercase\expandafter{\romannumeral3}-B and \uppercase\expandafter{\romannumeral3}-C that the
energy requirements per sensor are always higher than $(1-\rho)\mu$ and always lower than
$(1+\rho)\mu$, which can be guaranteed by choosing the data dimensionality or the projections
dimensionality appropriately. It is clear that if the energy requirements per sensor are always
lower than $(1-\rho)\mu$ then the calculated probabilities are equal to one, whereas if the energy
requirements per sensor are always higher than $(1+\rho)\mu$ then such probabilities are equal to
zero.} the probability of incorrect data collection due to energy depletion can be lower bounded
as follows:
\begin{equation}\label{eq:prob_raw_data}
\begin{split}
\emph{P}_\texttt{raw}
&\geq1-\emph{P}\left(\xi_1\geq n\tau ,\xi_2\geq n\tau \right)\\
&=1-\left(\frac{(1+\rho)\mu-n\tau}{2\rho\mu}\right)^2.
\end{split}
\end{equation}

\subsection{CS Data Acquisition Scheme}
We now consider the probability of incorrect data reconstruction in a CS based data acquisition
scheme by assuming that the signals sensed by the two sensors exhibit the same sparsity level,
i.e. $s=s_1=s_2$. In particular, we use the fact that $m_k \approx
c\left(s_k,n\right)=\mathcal{O}\left(s_k \log\frac{n}{s_k}\right)$ where $m_k \leq n$ is a
necessary (and sufficient) condition for the successful reconstruction of the sparse signals via
the $\ell_1$-norm minimization problem in (\ref{eq:CS_L1})~\cite{donoho2006high}. Therefore, by
using the assumption that the EH processes follow independent uniform distributions in the
interval $[(1-\rho)\mu, (1+\rho)\mu]$, we can also lower bound the probability of incorrect data
reconstruction as follows:
\begin{equation}\label{eq:prob_CS}
\begin{split}
\emph{P}_\texttt{CS}&\geq1-\emph{P}\left(\xi_1\geq m_1\tau,\xi_2\geq m_2\tau\right)\\
&=1-\left(\frac{(1+\rho)\mu-c(s,n)\tau}{2\rho\mu}\right)^2.
\end{split}
\end{equation}


\subsection{DCS Data Acquisition Scheme}
We finally consider the probability of incorrect data reconstruction in a DCS based data
acquisition scheme by assuming that the signals sensed by the two sensors exhibit a certain common
support size $s_c'$ and the same innovation support size, i.e., $s'=s_1'=s_2'$. We now use the
fact that

\begin{equation}\label{eq:DCS_bound1}
m_k\approx  c\left(\hat{s},n\right) \qquad k=1,2,
\end{equation}
where $\hat{s}=2s'-\frac{s'^2}{n}$ and
\begin{equation}\label{eq:DCS_bound2}
m_1+m_2\approx c\left(\hat{s},n\right)+c\left(\tilde{s},n\right)
\end{equation}
where $\tilde{s}=s_c'+2s'-\frac{2s_c's'}{n}-\frac{s'^2}{n}+\frac{s_c's'^2}{n^2}$, are necessary
conditions for the joint successful reconstruction of the two sparse signals via the joint
$\ell_1$ reconstruction algorithm in (\ref{eq:CS_DCS})~\cite{baron2006distributed}. Therefore, we
can also lower bound the probability of incorrect data reconstruction by using the assumption that
the EH processes follow independent uniform distributions in the interval $[(1-\rho)\mu,
(1+\rho)\mu]$ as follows:
\begin{equation}\label{eq:prob_DCS}
\begin{split}
&\emph{P}_\texttt{DCS} \geq1-\emph{P}\left(\xi_1\geq  c\left(\hat{s},n\right)\tau,\ \xi_2\geq c\left(\hat{s},n\right)\tau, \right. \\
&\qquad\qquad\qquad \left. \xi_1+\xi_2\geq
c\left(\hat{s},n\right)\tau+ c\left(\tilde{s},n\right)\tau\right)\\
& \geq1-\emph{P}\left(\xi_1+\xi_2\geq
c\left(\hat{s},n\right)\tau+ c\left(\tilde{s},n\right)\tau\right)\\
&=\left\{
\begin{array}{cl}
\frac{1}{2}\left(\frac{m' \tau-2(1-\rho)\mu}{2\rho\mu}\right)^2&\quad
\texttt{if}\ m'\tau\leq 2\mu\\
1-\frac{1}{2}\left(\frac{2(1+\rho)\mu-m' \tau}{2\rho\mu}\right)^2&\quad \texttt{if}\
m'\tau>2\mu
\end{array},
\right.
\end{split}
\end{equation}
where $m'=c\left(\hat{s},n\right)+c\left(\tilde{s},n\right)$.


\begin{figure}[!t]%
\centering%
\includegraphics[width=0.5\textwidth]{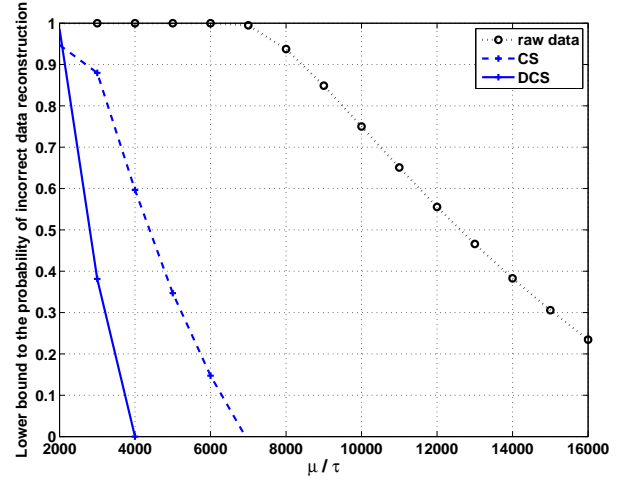}%
\DeclareGraphicsExtensions. \caption{Lower Bound to the probability of incorrect reconstruction
vs. $\mu/\tau$ ($\rho=0.5$, $n=10^4$, $s=10^3$, $s'=200$ and $s_c'=800$).} \label{fig:prob_bound}
\end{figure}%
\par
The fact that the lower bounds to the probability of incorrect data reconstruction for the raw
data gathering scheme in (\ref{eq:prob_raw_data}) or the CS data gathering scheme in
(\ref{eq:prob_CS}) are higher than the lower bound to the probability of incorrect data
reconstruction for the DCS data acquisition scheme in (\ref{eq:prob_DCS}) suggests that our
proposed approach leads to better reconstruction quality. These probability bounds are compared in
Fig.~\ref{fig:prob_bound} by using the approximation to the overmeasuring factor
$c\left(s_k,n\right)=s_k\log_2\left(1+\frac{n}{s_k}\right)$ in~\cite{baron2006distributed}.

\subsection{Towards Energy Neutrality: Matching Energy Demand to the Energy Supply}
The essence of the mechanism that offer the means to match the energy demand to the energy supply
at each SN in order to unlock energy neutrality is embodied in (\ref{eq:DCS_bound1}) and
(\ref{eq:DCS_bound2}): it is clear that -- together with the fact that the value of right hand
side of (\ref{eq:DCS_bound1}) is lower than the value of the right hand side of
(\ref{eq:DCS_bound2}) -- one can strike a trade-off between the number of measurements taken by
the two SNs in order to adapt the energy consumption to the energy availability at any particular
node without compromising the data reconstruction quality. That is, such a ``sensing diversity"
mechanism which is linked to the DCS data gathering process is key to adapt to the ``energy
diversity" mechanism which is linked to the EH process in order to guarantee successful data
recovery.

\par
It is also relevant to reflect on how some of the results may generalize with the increase in the
number of SNs in the network. It is clear that the probability that energy supply is not
sufficient for the energy demand increases substantially with the increase in the number of SNs in
the WSN. Hence, the probability of incorrect data reconstruction associated with the CS based data
acquisition scheme (and the raw one) also increases with the increase in the number of nodes. In
contrast, the probability of incorrect data reconstruction associated with the DCS based data
acquisition scheme exhibits a completely different behavior: such a probability decreases with the
increase in the number of nodes until a certain optimum value of SNs; and it then increases with
the increase in the number of nodes past the optimum value of SNs.
\par
The roots for such a behavior are linked to the interplay between the level of energy diversity
and the level of sensing diversity. For a number of SNs less than the optimal value of SNs the
common component in the SCCI model guarantees that there is enough "sensing diversity" in order to
match the energy demand to the variability of the energy supply. For a number of SNs higher than
the optimal value of SNs the innovations component in the SCCI model gradually compromises the
level of "sensing diversity". The exact optimal value of SNs then depends on the exact levels of
sparsity in the common and innovations components of the adopted model.
\par
Such a behavior is crisply illustrated in the sequel by considering experiments both with
synthetic and real data.


\section{Experimental Study}
We now illustrate the potential of the approach both with synthetic data and real data collected
by the WSN located in the Intel Berkeley Research Lab~\cite{intellabWSN}.

\subsection{Synthetic Experiments}
In the synthetic experiments, we generate the sparse signal representations $\mathbf{x}_k$
($k=1,\ldots,K$) randomly with ambient dimension $n=50$ are generated randomly, where the non-zero
components are drawn from independent identically distributed (i.i.d.) Gaussian distribution
$\mathcal{N}(0,1)$. We also generate the equivalent sensing matrices $\mathbf{A}_k$
($k=1,\ldots,K$) randomly, where the elements are also drawn from i.i.d. Gaussian distribution
$\mathcal{N}(0,1)$.

\begin{figure}[!t]%
\centering%
\includegraphics[width=0.5\textwidth]{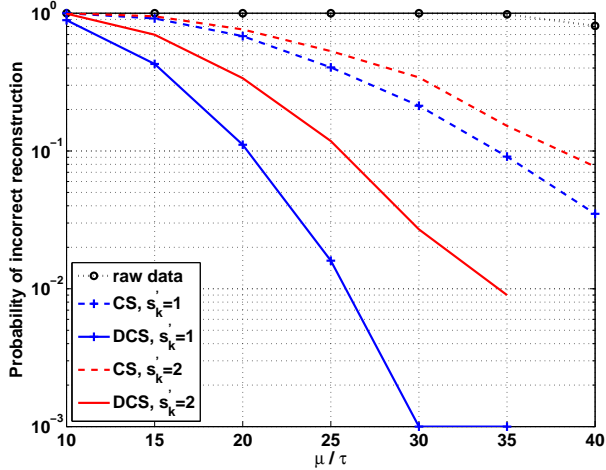}%
\DeclareGraphicsExtensions. \caption{Probability of incorrect reconstruction vs. $\mu/\tau$ ($K=2$
and $s_c'=5$).} \label{fig:prob_uni}
\end{figure}%
\par
Fig.~\ref{fig:prob_uni} shows the probability of incorrect data reconstruction vs. the energy
availability level for the different data gathering mechanisms. We can conclude that for a certain
target probability of incorrect data reconstruction the DCS scheme requires less EH capability in
relation to the CS or the raw data scheme or -- conversely -- for a certain EH capability the
proposed scheme leads to a lower probability of incorrect data reconstruction in relation to the
other schemes. For example, for probability of incorrect reconstruction equal to $0.01$, DCS with
$s_k=1$ requires only $26\tau$ while CS require $34\tau$, and raw data transmission requires
orders of magnitude higher than DCS in terms of mean energy availability. We can also conclude
that -- as argued in the previous section - that the sparsity level of the innovations component
of the SCCI model has a considerable effect on the trends and results.

\begin{figure}[!t]%
\centering%
\includegraphics[width=0.5\textwidth]{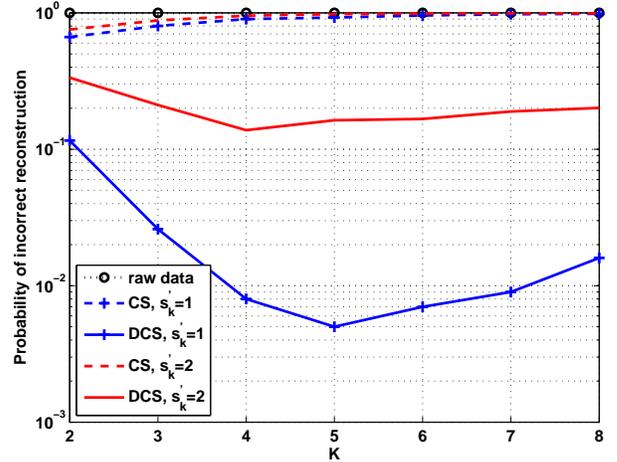}%
\DeclareGraphicsExtensions. \caption{Probability of incorrect reconstruction vs. number of SNs $K$
($\mu=20\tau$, $s_c'=4$ and $s_k'=1$).} \label{fig:prob_uni_k}
\end{figure}%

\par
Fig.~\ref{fig:prob_uni_k} shows the probability of incorrect data reconstruction vs. the number of
SNs in the EH WSN. This Figure confirms that the probability of incorrect reconstruction with CS
increases as the number of SNs grows, since a larger number of SNs yields a higher risk that some
SN will fail to harvest enough energy for acquiring the necessary number of measurements for
successful reconstruction. Conversely, the probability of incorrect reconstruction with the DCS
approach first decreases and then increase as the the number of SNs grows. This trend -- as argued
earlier -- depends on the interplay between the level of energy diversity and the level of sensing
diversity.

\subsection{Experiments With Real Data}
In the real data experiments, we illustrate the potential of the paradigm in a simple scenario:
temperature monitoring by a WSN located in the Intel Berkeley Research lab~\cite{intellabWSN}. In
particular, we only use the contiguous data that was available from $8$ SNs, i.e., SN 2, 3, 4, 7,
8, 9, 10 and 11. We assume the use of a typical 250 $\texttt{kbps}$ 62.64 $\texttt{mW}$
($17.4\texttt{mA} \times 3.6\texttt{V}$) ZigBee RF transceiver and a solar panel with an average
harvesting power 10$\mu \texttt{W/cm}^2$~\cite{roundy2004power}\footnote{Note that we ignore the
sensing energy cost in this investigation as transmission energy is much higher than the energy
cost in compressive non-uniform random sampling~\cite{6155205}.}. We also assume that harvested
power is uniformly distributed within $[5\mu \texttt{W/cm}^2, 15\mu \texttt{W/cm}^2]$. The SN
independently and randomly collects a small portion of the original samples and transmits them to
the FC based on their availability of harvested energy. All the temperature signals we employ in
the following study have a length of $n=512$.

\begin{figure}[!tb]%
\centering%
\includegraphics[width=0.5\textwidth]{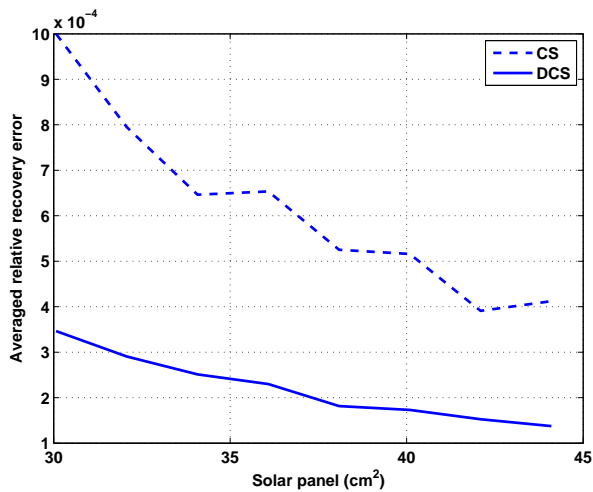}%
\DeclareGraphicsExtensions. \caption{Averaged relative recovery error vs. solar panel size ($K=2$).}
\label{fig:mse_uni}
\end{figure}%

\begin{figure}[!tb]%
\centering%
\includegraphics[width=0.5\textwidth]{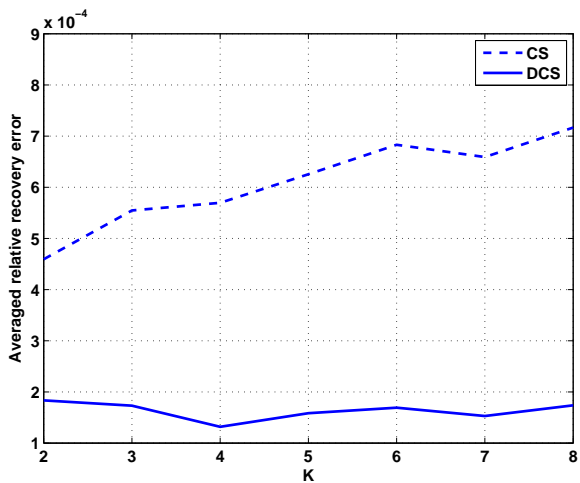}%
\DeclareGraphicsExtensions. \caption{Averaged relative recovery error vs. number of SNs $K$ (solar
panel of 40$cm^2$ ).} \label{fig:mse_uni_k}
\end{figure}%
\par
Note that the temperature signals monitored by the WSN are compressible rather than exactly sparse
via the discrete cosine transform (DCT). Thus, we use the relative recovery error for a single SN
which is equal to $\frac{\|\hat{\mathbf{f}}_k-\mathbf{f}_k\|_{2}^2}{\|\mathbf{f}_k\|_{2}^2}$,
where $\mathbf{f}_k$ and $\hat{\mathbf{f}}_k$ are the original signal and the reconstructed signal
of the $k$th SN respectively rather than the probability of incorrect data reconstruction to
evaluate the performance.

\par
Fig.~\ref{fig:mse_uni} shows the averaged relative recovery error of $K=2$ SNs, i.e., SN 2 and SN
3, achieved by the DCS and the CS data gathering schemes for various solar panel sizes. It is
clear that the DCS scheme requires much lower energy levels in relation to the CS scheme for a
certain target relative recovery error. For example, for an averaged relative recovery error equal
to $3\times 10^{-4}$, DCS requires only $32\ \texttt{cm}^2$ solar panel while CS requires a solar
panel exceeding $44 \texttt{cm}^2$ .

\par
Fig.~\ref{fig:mse_uni_k} shows the averaged relative recovery error with a solar panel with fixed
size achieved by the DCS and the CS schemes for various number of SNs. We note -- once again --
that the CS scheme fails as the number of SNs increases, but the DCS scheme does not.

\par
To conclude, the settings behind Figs.~\ref{fig:mse_uni} and~\ref{fig:mse_uni_k} are such that the
WSN is powered only via the energy harvested from the environment: the fact that DCS based data
gathering enables one to collect data without penalties on the data reconstruction error forms the
basis of the energy neutrality claims.

\section{Conclusion and Discussion}
It has been established that energy diversity and sensing diversity are two fundamental mechanisms
that offer the means to match the energy demand to the energy supply in EH WSNs based on the DCS
data acquisition paradigm. It has also been established that DCS based data acquisition paradigm
provides substantial gains in energy efficiency for a certain target data reconstruction quality
in relation to other approach, e.g. CS based data acquisition.

\par
The potential of the approach has been unveiled for a simple energy management approach, where the
SNs choose to use the entire energy budget per activation interval rather than use only a fraction
of the energy budget and store the remaining fraction in some local battery. It is clear that such
a more refined energy management approach will also yield to further gains.

\par
Finally, it is interesting to point out that the underlying diversity principles are not too
dissimilar from the diversity principles in wireless communications: the variability of the
wireless channel calls for transmission methods capable of providing robustness; in turn, the
variability of the energy harvesting calls instead for robust sensing methods.


%


%
%



\bibliographystyle{IEEEtran}
\bibliography{IEEEabrv,GLOBECOM}

\end{document}